\documentclass[prb,citeautoscript,twocolumn,preprintnumbers,amsmath,amssymb,superscriptaddress]{revtex4}

\usepackage{graphicx}% Include figure files
\usepackage{dcolumn}% Align table columns on decimal point
\usepackage{bm}% bold math

\newcommand{\picwidth}{0.9\linewidth}
\newcommand{\q}[1]{``#1''}

\newcommand{\ub}{U_{\text{B}}}
\newcommand{\ud}{U_{\Delta}}
\newcommand{\vd}{v_{\text{d}}}
\newcommand{\Eexc}{\text{E}_\mathrm{exc}}
\newcommand{\muexc}{\mu_\mathrm{exc}}

\newcommand{\PL}{PL}
\newcommand{\TOF}{TOF}

\newcommand{\cmVs}{cm$^2$/eVs}

\begin{document}

\title{Drift mobility of long-living excitons in coupled GaAs quantum wells}

\author{A. G\"artner}
\author{A. W. Holleitner}
\email[Author to whom correspondence should be addressed;
\mbox{E-Mail:\ }]{Alex.Holleitner@physik.uni-muenchen.de}
\affiliation{
Department f\"ur Physik and Center for NanoScience,
Ludwig-Maximilians-Universit\"at,
Geschwister-Scholl-Platz 1, D-80539 M\"unchen, Germany
}
\author{D. Schuh}
\affiliation{
Institut f\"ur Angewandte und Experimentelle Physik,
Universit\"at Regensburg, Universit\"atsstra{\ss}e 31,
D-93040 Regensburg, Germany
}
\altaffiliation{
former address: 
Walter Schottky Institut, Technische Universit\"at M\"unchen, 
Am Coulombwall 3, D-85748 Garching, Germany
}
\author{J. P. Kotthaus}
\affiliation{
Department f\"ur Physik and Center for NanoScience,
Ludwig-Maximilians-Universit\"at,
Geschwister-Scholl-Platz 1, D-80539 M\"unchen, Germany
}

\begin{abstract}

We observe high-mobility transport of indirect excitons in coupled
GaAs quantum wells. A voltage-tunable in-plane potential gradient
is defined for excitons by exploiting the quantum confined Stark
effect in combination with a lithographically designed resistive
top gate. Excitonic photoluminescence resolved in space, energy,
and time provides insight into the in-plane drift dynamics. Across
several hundreds of microns an excitonic mobility of
$>10^5$\,\cmVs\ is observed for temperatures below $10$\,K. With
increasing temperature the excitonic mobility decreases due to
exciton-phonon scattering.

\end{abstract}
\pacs{}
\keywords{Excitonic Transport, Long-living Indirect Exciton,
Mobility, Coupled Quantum Well}

\maketitle

The pioneering work of Keldysh and Kozlov in 1968 has triggered
many experiments aiming to observe the bosonic nature of excitons
in solid state systems~\cite{KelJETP68}. For detecting the
Bose-Einstein condensation of excitons, it is a prerequisite to
define controllable confinement potentials for excitons. So far
trapping of excitons has been demonstrated in strained systems
 \cite{Trauernicht83, Kash88, SnoAPL99}, magnetic traps
 \cite{Christianen98}, \q{natural traps} defined by interface
roughness fluctuations \cite{ButNat02a}, and electrostatic traps
\cite{ZimAPL98, Huber98, KraPRL02, Ham05}. Only the latter enable
in-situ control of the trapping potential. In addition,
electrostatic traps can be extended towards optoelectronic
solid-state devices due to their potential scalability and
compatibility with existing semiconductor technology.
\newline
\begin{figure}[t]
\includegraphics[width=\picwidth]{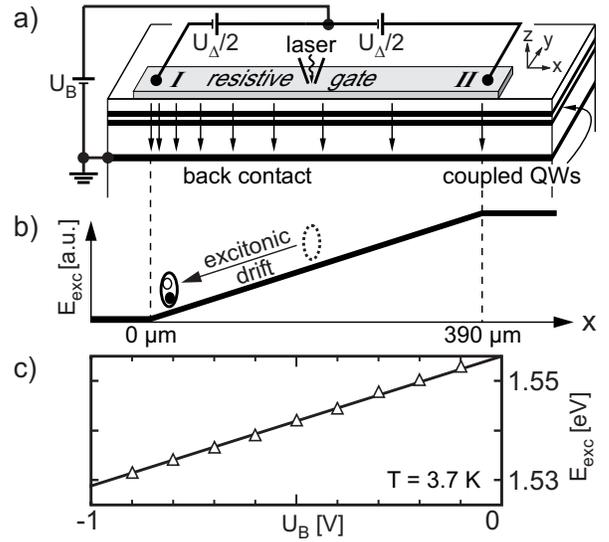}
\caption{
   \label{fig1}
   (a) Excitonic time-of-flight apparatus. A current-carrying
   top gate (grey) defines a laterally varying vertical electric
   field (vertical arrows). (b) Sketch of the in-plane excitonic
   potential between contacts \q{I} and \q{II} due
   to the quantum confined Stark effect (QCSE).
   The slope of the gradient is tunable via the voltage $\ud$.
   (c) Calibration of the QCSE-shift by application of a
   dc-voltage $\ub$ to the top gate.
   }
\end{figure}
Here we investigate the drift dynamics of long-living excitons in
coupled quantum wells (QWs) in a voltage-tunable semiconductor
device. In prior experiments on coupled GaAs/AlAs-QWs a static,
spatially resolved photoluminescence spectroscopy has been used to
detect excitonic drift \cite{HagAPL95}. We extend this approach
towards time-of-flight (TOF) experiments in coupled GaAs-QWs by
detecting the excitonic photoluminescence (PL) as a function of
space, energy, and time. The technique relies on the quantum
confined Stark effect (QCSE), and it allows distinguishing the
dynamics of excitons from electron-hole effects \cite{KraPRL02}.
In a field effect
structure such as shown in Fig.~1(a), electrons and holes of
photogenerated excitons may rearrange in a way that they are
spatially separated by the tunnel barrier between the GaAs-QWs.
These indirect excitons have a lifetime of $\sim 300$\,ns (for
perpendicular electric fields of $\sim
10^6$\,V/m) \cite{GarPhysE06}, while the lifetimes of direct
excitons are in the order of $1$\,ns \cite{Feldmann}. The
excitonic drift of such indirect, long-living excitons is induced
by applying a voltage $\ud$ across a resistive top gate (Fig.
1(a)). Hereby, the electric field perpendicular to the QWs is
laterally varied, and due to the QCSE, mobile excitons drift along
the gradient towards regions of high electric field (Fig.
1(b)) \cite{ZimAPL98, HagAPL95}. In the TOF-experiments we find
excitonic mobilities $\geq 10^5$\,\cmVs\ and scattering times
$\geq 10$~ps at low temperatures. Both values exceed previous
results on coupled QWs by a factor of $200$ \cite{HagAPL95}. For
temperatures higher than $10$~K the excitonic drift is limited by
phonon-scattering processes.
\newline
\indent Starting point is an epitaxially grown
AlGaAs/GaAs-heterostructure containing two GaAs-QWs encompassed by
AlGaAs barriers~(Fig.~\ref{fig1}(a)). Each QW has a thickness of
8\,nm, while the QWs are separated by a 4\,nm-thick tunnel barrier
made out of Al$_{0.3}$Ga$_{0.7}$As. The QWs are located 60\,nm
below the surface of the heterostructure. An n-doped GaAs-layer at
a depth of $d=370$\,nm serves as back gate, and a semi-transparent
titanium layer is used as the top gate of the field effect
structure. The metal gates, prepared by standard optical
lithography, have typically a thickness of 10\,nm, a width of
50~$\mu$m, and a length ranging between $500$\,$\mu$m and
$1000$\,$\mu$m. The resistance of such a gate strip is between
$2$\,k$\Omega$ and $4$\,k$\Omega$ depending on its length.
\newline
\indent The excitonic drift experiments are carried out in a
helium continuous-flow cryostat in combination with a
micro-photoluminescence setup in the temperature range between
3.5\,K and 40\,K. The excitons are locally excited by focusing a
pulsed laser onto the center of the top gate. The laser is 
operated at a pulse length of $50$\,ns and at a repetition 
period of $10$\,$\mu$s. At a spot diameter of $\approx 10$\,$\mu$m the
power density is 5\,kW/cm$^2$. The laser wavelength is chosen to
be 680\,nm, such that electron-hole-pairs are only created in the
GaAs-QWs and not in the AlGaAs-barriers. For the \TOF\
experiments, the \PL\ signal of the recombining excitons is picked
up by the optical microscope as a function of the time-delay with
respect to the initial laser pulse. The optical signal is
subsequently guided through a triple-grating imaging spectrometer.
An attached fast-gated, intensified CCD (charge coupled device)
camera with an exposure time of 5\,ns detects the \PL\ emission of
the excitonic cloud as a function of energy and space. In order to
yield a sufficient signal to noise ratio, all images shown are
taken by integrating over $2 \times 10^7$ single events.
\newline
\indent In Fig.~1(c) we calibrate the shift of the exciton energy
due to the QCSE as a function of the applied voltage $\ub$ at $\ud=0$\,V.
The energy $\Eexc$ of the spatially indirect excitons is shifted to
lower values by $\delta \Eexc = ed \times E_z$, with $d$ the
center distance of the two QWs and $E_z$ the electric field
perpendicular to the QWs. The data in Fig.~1(c) nicely follow a
linear dependence with a slope of $\partial \Eexc /
\partial U = 26.4$\,meV/V. The red-shift is independent of the bath
temperature $T$ up to 30\,K. For the \TOF\ experiments, a constant
bias voltage $\ub$ of $-0.4$\,V is applied to the top gate with
respect to the grounded back contact at all times. $50$\,ns after
the laser has been switched off, all short-living direct excitons have
decayed and only indirect excitons remain. Due to diffusion such a
cloud of mobile indirect excitons has typically a FWHM-diameter of
about $80$\,$\mu$m, in accordance with previous
results \cite{SnoPRL05, ButPRL05}.
\newline
\indent We define $t = 0$ as the point of time when the voltage
drop $\ud$ is applied across the resistive gate strip. In turn,
the voltage between contacts \q{I} and \q{II} increases linearly
along the gate strip. This voltage configuration creates a
QCSE-mediated excitonic gradient potential $\Eexc$ as sketched in
Fig.~\ref{fig1}(b). In turn, the excitons are exposed to a force
$F = -\nabla \Eexc = ed\nabla |E_z|$ \cite{GarPhysE06}. Since
$\ud$ is widely tunable, the method allows studying the in-plane
drift of indirect excitons at different velocities.
\begin{figure}[t]
\includegraphics[width=\picwidth]{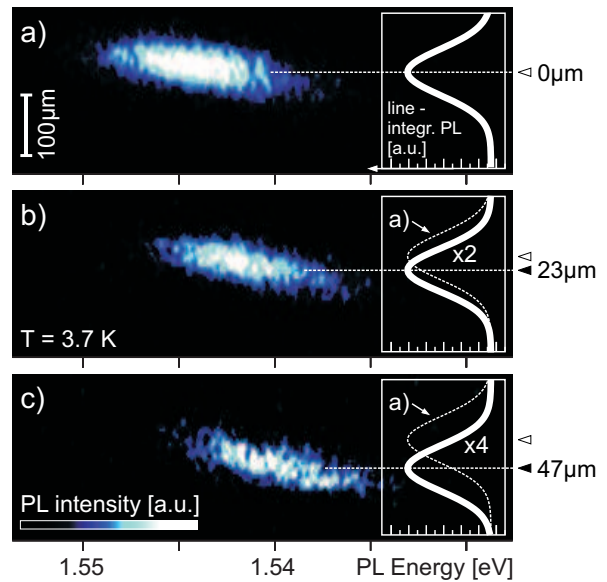}
\caption{
   \label{fig2}
   Photoluminescence images showing drift of excitons at $T = 3.7$\,K;
   taken at $t=0$\,ns (a), $t=20$\,ns (b), and $t=40$\,ns
   (c) after enabling the excitonic gradient potential.
   Insets:
   Energy-integrated representations of the data (solid curves).
   In the insets of (b) and (c) the lateral
   \PL\ distribution of the cloud at $t=0$\,ns is shown as a dashed curve.
}
\end{figure}
Fig.~\ref{fig2} shows three subsequent CCD-snapshots with 5\,ns
exposure time for $\ud = 2$\,V at (a) $t = 0$, (b) $t = 20$\,ns,
and (c) $t = 40$\,ns. Each image exhibits the lateral distribution of
the excitons resolved in space (vertical axis) and in energy
(horizontal axis). The distributions are tilted off the horizontal
orientation in all snapshots, proving that the gate strip creates
an in-plane gradient of the excitonic energy. Under the given
experimental conditions, the linear gradient $\nabla \Eexc \approx
100$\,$\mu$eV/$\mu$m obtained from the tilt in Fig.~2(b) and (c)
agrees well with the value for the dc-QCSE energy shift presented
in Fig.~1(c). The lower energy gradient in Fig.~2(a) is due to an
$RC$-constant of the resistive gate of $<10$\,ns which governs the
raising behaviour of $\ud$.
\newline
\indent In Fig.~2(b)[(c)] the center of the excitonic cloud has
travelled $(23.3 \pm 1.1)$\,$\mu$m [$(47.4 \pm 1.8)$\,$\mu$m] away
from the excitation spot towards electrode \q{I} as defined in
Fig.~1(a). At the same time, the excitons have reduced their
energy by $\sim$ 2\,meV [$\sim$ 4\,meV]. As a function of the
delay-time, the center of the excitonic distribution follows a
diagonal path with respect to space and energy within the error
bars (data not shown). The corresponding gradient $\partial \Eexc
/ \partial U\approx 25$\,meV/V again agrees well with the gradient
obtained from the dc-measurements in Fig.~1(c). This experimental
finding proves that we study the drift dynamics specifically of
indirect excitons. In addition, we would like to note that due to
the finite lifetime of the indirect excitons the \PL\ intensity in
Fig.~2(b) and 2(c) has decreased by a factor of $2$ and $4$,
respectively.
\newline
\indent By following the temporal evolution of the center of the
cloud in Fig.~2, the excitonic drift velocity $\vd$ can be
directly determined. Fig.~\ref{fig3}(a) shows the dependence of
$\vd$ on the voltage drop $\ud$ for various bath temperatures. At
low temperatures we observe a maximum velocity of about $\sim 2.5
\times 10^3$\,m/s. Experimentally, a leakage current $I_\mathrm{L}$
between the top and the back gate limits further increase of
$\ud$ and thus $\vd$. For $\ud \leq 2.5$\,V we find $I_\mathrm{L}$ to be $\leq
1$\,$\mu$A. In this regime, a linear fit of the data gives the
differential mobility $\muexc$ of the excitons defined as
\begin{equation}
\label{mu}
{  \mu_\mathrm{exc} = L \cdot (\mathrm{d} \vd}/{\mathrm{d}\ud})\cdot
(-{\partial \Eexc}/{\partial U})^{-1}\ .
\end{equation}
\begin{figure}[t]
\includegraphics[width=\picwidth]{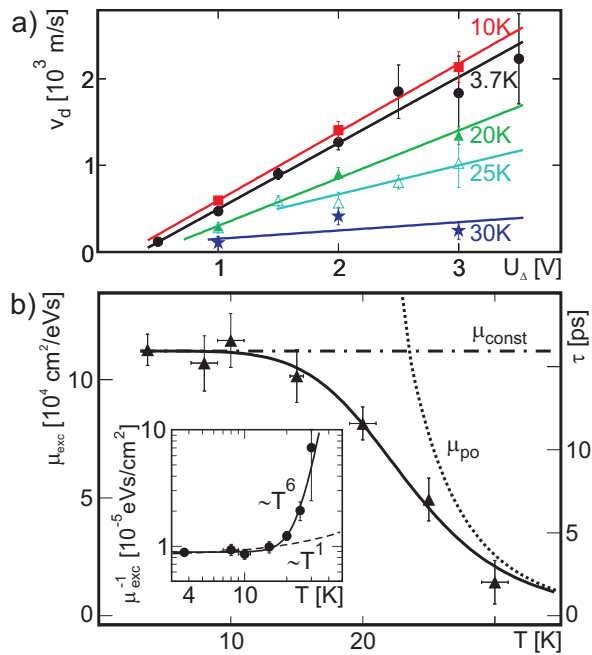}
\caption{
   \label{fig3}
   (a) Measured excitonic drift velocity $\vd$ versus the voltage
    drop $\ud$ at various bath temperatures $T$. The differential
    mobility is obtained from linear curves fitted to the data.
   (b) Excitonic mobility $\muexc$ and scattering time $\tau$
   (right vertical axis) as a function of the temperature.
   Inset: double logarithmic representation of the inverse mobility
   $1/\muexc$ versus temperature. Solid lines are theoretical
   fits to the data.
}
\end{figure}
Fig.~\ref{fig3}(b) summarizes the differential mobility
$\mu_\mathrm{exc}$ of indirect excitons at various temperatures.
At low temperature, a constant mobility $\mu_{\mathrm{const}}$ larger than
$10^5$\,\cmVs\ is observed, which exceeds previous results by a
factor $\geq 200$ \cite{HagAPL95}. Recent publications on GaAs-QWs
suggest that the maximum mobility in our experiment is only
limited by barrier alloy scattering \cite{Hillmer89}, which is
independent of temperature (dashed dotted line). With increasing
temperature the excitonic mobility decreases as $\mu_\mathrm{po} \propto
T^{-6}$ (dotted line), which is usually referred to enhanced
scattering of excitons by polar optical phonons \cite{Basu91}. The
combination of both scattering processes according to
Matthiessen's rule explains the experimental data well
(solid line). We would like to note that polar optical phonon scattering
is usually expected for temperatures above 100\,K \cite{Basu91}.
However, as seen in the inset of Fig.~3(b), acoustic phonon
scattering ($\propto T^{-1}$) cannot explain the data (dashed
line) \cite{Hillmer89}. Assuming an effective exciton mass
$m^*_\mathrm{exc} = (m^*_\mathrm{e} + m^*_\mathrm{h}) \approx 0.25$\,$m_\mathrm{e}$ with
$m^*_\mathrm{e,h}$ the effective electron/hole mass in GaAs and
$m_\mathrm{e}$ the free electron mass, we can further estimate the
transport scattering time to be $\tau \equiv \muexc \cdot
m^*_\mathrm{exc}$ (right axis in Fig.~\ref{fig3}(b)). Since the
\PL\ signal depends on the temperature as well, \TOF\ experiments
above $30$\,K are ambiguous. Generally, the data shown are obtained
from different samples patterned on one AlGaAs/GaAs-wafer. Since
the electron mobility is proportional to the sixth power of the QW
width \cite{Sasaki}, future experiments will aim towards wider
QWs.
\newline
\indent In summary, we explore the quantum confined Stark effect
in combination with a resistive top gate to study drift dynamics
of indirect excitons in coupled GaAs quantum wells. The emitted
photoluminescence of the drifting excitons is resolved in space,
energy, and time, which allows measuring the drift velocity and
mobility of the excitons. At low temperatures we observe a maximum
mobility of $10^5$\,\cmVs\ which is a factor of 200 times larger
than previous results on long-living excitons in coupled quantum
wells.
\newline
\indent We thank S. Manus and L. Prechtel for technical assistance
and the Deutsche Forschungsgemeinschaft for financial support.
%%%
%%%
%%%
%%%%%%%%%%%%%%%%%%%%%%%%%%%%%%%%%%%%%%%%%%%%%%%%%%%%%%%%%%%%%%%%%%
%\section*{\label{sec:ref}References}
%%%%%%%%%%%%%%%%%%%%%%%%%%%%%%%%%%%%%%%%%%%%%%%%%%%%%%%%%%%%%%%%%%
%

\end{document}